\documentclass[8pt,a4paper,twocolumn,english]{extarticle}
\usepackage[T1]{fontenc}
\usepackage[latin9]{inputenc}
\usepackage{textcomp}
\usepackage{graphicx}
\usepackage{subscript}

\makeatletter

\newcommand{\binom}[2]{{#1 \choose #2}}

\providecommand{\tabularnewline}{\\}

\usepackage{a4wide}

\makeatother

\usepackage{babel}
\begin{document}

\title{Spatiotemporal model of a key step in endocytosis: \\
SNX9 recruitment via phosphoinositides}

\author{Johannes Schöneberg$^{\dagger1}$, Alexander Ullrich$^{\dagger1}$,
York Posor$^{\ddagger}$, Volker Haucke$^{\ddagger}$, Frank Noe$^{\dagger*}$}
\maketitle
\begin{abstract}
Clathrin mediated endocytosis (CME) is an ubiquitous cellular pathway
that regulates central aspects of cell physiology such as nutrient
uptake, modulation of signal transduction, synaptic transmission and
membrane turn-over. Endocytic vesicle formation depends on the timed
production of specific phosphoinositides and their interactions with
various endocytic proteins. Recently, it has been found that phosphatidylinositol-3,4-bisphosphate
({\normalsize{$\mathrm{PI}(3,4)\mathrm{P}_{2}$)}} produced by the
class II phosphatidylinositol 3-kinase C2$\alpha$ plays a key role
in the recruitment of the PX-BAR domain protein SNX9, which is proposed
to play a role in the constriction of the endocytic vesicle neck
\cite{Posor2013}. {\normalsize{Interestingly, SNX9 and its close
paralog SNX18 are not fully specific to $\mathrm{PI}(3,4)\mathrm{P}_{2}$
but can also bind other phospholipids, in particular to}} {\normalsize{$\mathrm{PI}(4,5)\mathrm{P}_{2}$,
an abundant plasma membrane lipid required for the recruitment of
many endocytic proteins.}} In order to understand the dynamical interplay
between phospholipids and endocytic proteins, we developed a computational
model of the temporal changes in the population of the phosphoinositide-associated
endocytic proteins and their spatial distribution at a clathrin-coated
pit (CCP). The model resolves single molecules in time and space,
and incorporates the complex interplay of proteins and lipids, as
well as their movement within the CCP. We find that the comparably
small differences in lipid binding affinities of endocytic proteins
are amplified by competition among them, allowing for the selective
enrichment of SNX9 at late stage CCPs as a result of timed {\normalsize{$\mathrm{PI}(3,4)\mathrm{P}_{2}$}}
production. 
\end{abstract}
\vspace{0.5cm}

$\dagger$: FU Berlin, Department of Mathematics and Computer Science,
Arnimallee 6, 14195 Berlin, Germany

$\ddagger$: Leibniz Institut für Molekulare Pharmakologie, Robert-Roessle-Strasse
10, 13125 Berlin, Germany

1: equal contribution

{*}: correspondance to frank.noe@fu-berlin.de

\section*{Introduction}

Clathrin mediated endocytosis (CME) is a key process by which cells
internalize surface proteins including nutrient and signalling receptors,
adhesion proteins, or ion channels. CME can be divided into four steps,
CCP nucleation at the plasma membrane, CCP invagination, dynamin-mediated
scission of late stage CCPs and clathrin uncoating of the free vesicle.
While many key players and interactions of this endocytic pathway
have been studied intensively, we are just starting to understand
the spatiotemporal regulation of CME. An important factor in CME
are phosphoinositides and their timed formation by specific kinases
and phosphatases \cite{Zoncu2007,Loerke2009,Krauss2006}. Phosphatidylinositol-4,5-bisphosphate
{[}$\mathrm{PI}(4,5)\mathrm{P}_{2}${]} is known to play an important
role in CME by associating with CCP components such as adaptor protein
AP-2 and membrane curvature inducing proteins including epsins \cite{Henne2010a,Cocucci2012}.
Less is known about phosphoinositides other than $\mathrm{PI}(4,5)\mathrm{P}_{2}$
and their potential roles in regulating CME. The recent identification
of PI3K C2$\alpha$ and its lipid product phosphatidylinositol-3,4-bisphosphate
($\mathrm{PI}(3,4)\mathrm{P}_{2}$) as regulators of CME suggests
an interplay of two distinct PI-bisphosphate species, $\mathrm{PI}(4,5)\mathrm{P}_{2}$
and $\mathrm{PI}(3,4)\mathrm{P}_{2}$, in driving CME \cite{Posor2013}.
Formation of $\mathrm{PI}(3,4)\mathrm{P}_{2}$ appears to be spatiotemporally
coupled to $\mathrm{PI}(4,5)\mathrm{P}_{2}$-mediated CCP assembly
as recruitment of PI3K C2$\alpha$ depends on clathrin, which via
early acting adaptors FCHo, AP-2, or CALM selectively associates with
$\mathrm{PI}(4,5)\mathrm{P}_{2}$-rich membrane sites. Elevated levels
of $\mathrm{PI}(3,4)\mathrm{P}_{2}$ might then facilitate the enrichment
of $\mathrm{PI}(3,4)\mathrm{P}_{2}$-binding effector proteins, most
notably SNX9 prior to dynamin-mediated endocytic vesicle fission.
A prediction from these findings is that SNX9 is recruited to CCPs
subsequent to PI3K C2$\alpha$ but preceding the burst of dynamin
2 recruitment. TIRF microscopy analysis of the timing of endocytic
protein recruitment indeed revealed that accumulation of mCherry-SNX9
was delayed by about 20 s with respect to eGFP-PI3K C2$\alpha$ but
clearly preceded the arrival of the majority of dynamin 2 \cite{Posor2013},
consistent with a prominent role for $\mathrm{PI}(3,4)\mathrm{P}_{2}$
in stabilizing SNX9 at endocytic sites. This time course of SNX9 arrival
at CCPs coincides with a critical phase of the endocytic process during
which extensive membrane remodelling must occur as a narrow stalk
is formed that becomes a substrate for membrane fission by dynamin
(and possibly other proteins such as epsins\cite{Boucrot2012}). SNX9,
in spite of its apparent dependence on $\mathrm{PI}(3,4)\mathrm{P}_{2}$
production by PI3K C2$\alpha$ did not display a strong preference
for $\mathrm{PI}(3,4)\mathrm{P}_{2}$ over $\mathrm{PI}(4,5)\mathrm{P}_{2}$
\emph{in vitro}. Since different endocytic proteins may compete for
the same population of PIs in vivo, the protein-lipid population dynamics
may be complex and cannot be predicted easily based on single binding
affinity values. 

In this study, we designed a theoretical reaction-diffusion model
with single-molecule resolution, in order to grasp this complex interplay
of proteins and lipids. Computer simulations of this model can trace
population changes and spatial distribution of PI-binding proteins
at CCPs depending on PI concentration. The model is parametrized to
have a steady state with absolute protein copy numbers determined
from proteomic studies\cite{Borner2012} and uses published lipid
binding affinities of the major PI-binding endocytic proteins\cite{Henne2010a,Ford2001,Ford2002,Jackson2010}
(see Methods and \cite{Posor2013} for details).

\section*{Methods}

\subsection*{Single-particle reaction diffusion simulation model }

In order to model the distribution and diffusion of phospholipids
and endocytic proteins underneath the clathrin coat, a single-particle
reaction-diffusion model of a single CCP was developed. Computer simulations
of this model can trace phospholipid-dependent population changes
at CCPs depending on PI concentration. The model is parametrized to
have a steady state with absolute protein copy numbers determined
from proteomic studies and uses published lipid binding affinities
of the major PI-binding endocytic proteins (see Table S\ref{tab1:parameter}).
This model represents space in a coarse-grained manner, such that
it is not able to resolve fine details such as protein crowding as
nonuniform protein distributions reliably, but is rather suited to
efficiently calculate the time-dependent population changes of proteins
at the CCP. The simulation space consists of a hexagonal grid of
217 discrete simulation cells, representing a membrane patch of about
200 nm diameter (see Fig. \ref{fig1_model}a). In order to model a
prototypical clathrin cage, a fixed arrangement of 54 intertwined
clathrin triskelia was mapped on the grid (Fig. \ref{fig1_model}a).
Each clathrin triskelion exposes three binding sites to clathrin-binding
endocytic proteins underneath (Fig. \ref{fig1_model}b). In the simulation,
the clathrin coat has the sole purpose of defining the positions of
these binding sites. At any time, each simulation cell maintains the
following pieces of information: 
\begin{itemize}
\item Copy number of $\mathrm{PI}(3,4)\mathrm{P}_{2}$ and $\mathrm{PI}(4,5)\mathrm{P}_{2}$
molecules
\item Copy number of membrane-associated proteins for each of the following
proteins: SNX9, AP-2, FCHo2, CALM, Epsin. 
\item Number of clathrin binding sites, and which proteins are bound there,
if any. Each protein copy in the simulation maintains the following
data: 
\item The number of $\mathrm{PI}(3,4)\mathrm{P}_{2}$ or $\mathrm{PI}(4,5)\mathrm{P}_{2}$
molecules bound. 
\end{itemize}
The simulation proceeds in discrete time steps of 0.01 ms. In each
time step, all lipid molecules and all membrane-associated protein
molecules can diffuse laterally, and can undergo \textquotedblleft{}reactions\textquotedblright{},
corresponding to binding or unbinding events.

\subsubsection*{Diffusion: }

Both phosphoinositides and membrane-associated proteins can diffuse
laterally. Since each single copy of phosphoinositide and protein
is represented as an explicit object, diffusion corresponds to jump
events of lipids or proteins between neighboring simulation cells.
The probability with which a jump may be attempted within a simulation
time step is calculated from the diffusion constant, the simulation
cell surface area, and the time step duration. For lipids, all jump
attempts are executed. For proteins, the success of a jump event depends
on the size of the protein and the occupancy of the target simulation
cell. In this way, crowding effects are included in the model - proteins
can only diffuse to places where sufficient physical space is vacant.
Proteins do not diffuse while they are bound to clathrin. Any lipids
bound to a protein are also not considered for independent diffusional
jumps, but will always move together with the associated protein.

\subsubsection*{Reactions: }

\begin{figure*}[!t]
\centering{}\includegraphics[width=1\textwidth]{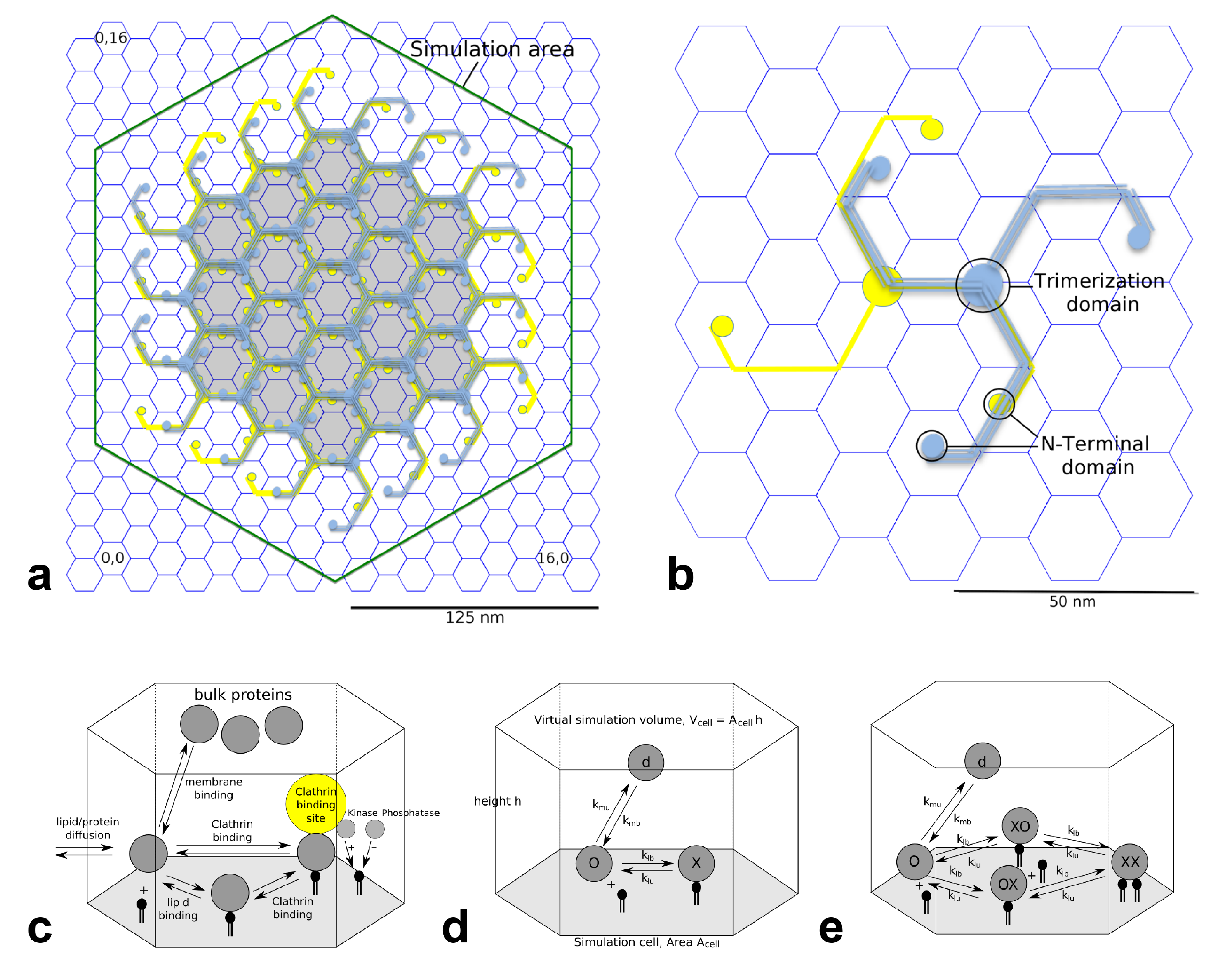}\caption{\label{fig1_model}Illustration of the single-particle reaction-diffusion
model of a CCP. a) Hexagonal simulation grid. The large green hexagon
enscribes the area simulated, consisting of 217 simulation cells (small
blue hexagons). The yellow and blue structures represent 54 clathrin
triskelia, with clathrin hubs shown as larger circles and binding
sites shown as smaller circles. The shaded gray area is considered
to be the area underneath the clathrin coat. b) Detailed view of two
nearby clathrin triskelia (blue and yellow). The central sphere represents
the hub of the clathrin triskelion, the terminal spheres represent
the clathrin binding sites available to endocytic proteins in the
same simulation cell. c) Summary of the reaction and binding processes
occuring in simulation cells. d,e) Sketch of binding model for proteins
with one/two phospholipid binding sites. Symbols: d: dissolved protein;
O: membrane bound protein, no lipid bound; X membrane bound protein,
one lipid bound; XO: membrane bound protein, lipid in left one of
two binding sites; OX: membrane bound protein, lipid in right one
of two binding sites; XX: membrane bound protein, lipids in both of
two binding sites; $k_{mb}$: protein-membrane binding rate; $k_{ub}$:
protein-membrane unbinding rate; $k_{lb}$: protein-lipid binding
rate; $k_{lu}$: protein-lipid unbinding rate. See Methods section
and Table S\ref{tab1:parameter} for further details about the model
and its parameterization.}
\end{figure*}

Within each simulation cell, several types of binding or unbinding
reactions may occur between clathrin, the endocytic proteins and phosphoinositides
contained in the cell (Fig. \ref{fig1_model}c-e). Since the simulation
proceeds by discrete time steps, one reaction event of each type is
considered per time step and is executed with the probability p =
1-exp(-k dt), where k is the rate and dt is the time step duration. 
\begin{itemize}
\item Protein association: Endocytic proteins are recruited to the membrane
from a small section of the cytosol 20nm above the membrane with a
rate constant $k_{on,membrane}$ (Table S\ref{tab1:parameter}). The
number of proteins available per simulation cell for such a recruitment
attempt is fixed by the bulk concentration (Table S\ref{tab1:parameter}).
A recruitment event is only executed, if the protein size can still
be accommodated on the target cell. 
\item Protein dissociation: Membrane-associated proteins that are bound
to neither phosphoinositide nor clathrin are only associated with
their intrinsic affinity. Such a protein may dissociate with a dissociation
rate defined by the intrinsic microscopic binding affinity, $K_{a,mem}$,
(Table S\ref{tab1:parameter}) and the binding rate. Upon dissociation,
it is then removed from the simulation cell. The cytosolic population
remains constant throughout the simulation. Membrane-associated proteins
that bind at least one phosphoinositide or are bound to a clathrin
binding site are not considered for dissociation. Such proteins must
first loose their phosphoinositide and/or dissociate from clathrin
before being considered for dissociation. 
\item Lipid binding: Phosphoinositide binding sites of any membrane-associated
protein can bind free phosphoinositides present in the same simulation
cell with a rate $K_{on,lipid}$ (Table S\ref{tab1:parameter}). 
\item Lipid unbinding: Each binding site associated with a phosphoinositide
can unbind this lipid with dissociation rates based on the binding
rates and the affinities $K_{a,lip45}$, $K_{a,lipweak45}$, $K_{a,lip34}$
and $K_{a,lipweak34}$ (Table S\ref{tab1:parameter}). 
\item Clathrin-Protein association: Proteins can bind to free clathrin binding
sites in the same simulation cell with a rate $K_{on,clathrin}$ (Table
S\ref{tab1:parameter}). Each clathrin binding site can only associate
with one protein at a time. The protein remains in the same simulation
cell but is removed from the freely diffusing pool. 
\item Clathrin-Protein dissociation: Proteins associated to clathrin binding
sites can dissociate with a rate according to the binding rate and
the affinity $K_{a,clathrin}$ (Table S\ref{tab1:parameter}). The
protein remains in the same simulation cell and is returned to the
freely diffusing pool. 
\item Production of $\mathrm{PI}(3,4)\mathrm{P}_{2}$ by PI3K C2$\alpha$:
$\mathrm{PI}(3,4)\mathrm{P}_{2}$ is produced by phosphorylation from
the implicit pool of PI(4)P. Since PI3K C2$\alpha$ is associated
with clathrin terminal domains, the reaction is conducted at clathrin
binding sites with a rate $k_{kinase}$ (Table S\ref{tab1:parameter})
whose value depends on the simulation parameter set used. Every such
reaction event increments the local $\mathrm{PI}(3,4)\mathrm{P}_{2}$
lipid count by one. The production of $\mathrm{PI}(3,4)\mathrm{P}_{2}$
is only started after a simulation time of 10 s. 
\item Depletion of $\mathrm{PI}(4,5)\mathrm{P}_{2}$ by phosphatases: In
some simulation parameter settings, we are considering the action
of phosphatases that deplete the $\mathrm{PI}(4,5)\mathrm{P}_{2}$
pool with a rate of $k{}_{phosphatase}$ (Table S\ref{tab1:parameter})
at every clathrin binding site. 
\end{itemize}
All reactions are described by a on- and an off-rate. Since no sufficient
experimental data was available to define all these rates, we extracted
the binding affinities (or dissociation constants) from experimental
data, and then set the absolute value of kon arbitrarily. The effect
of this arbitrary choice is that the kinetics, i.e. the speed at which
equilibria or steady states are reached in our simulation is not predictive
- in reality the equilibrium protein population could be reached faster
or slower than in our simulations after the change of the lipid concentration.
However, the equilibrium state itself, i.e. the number of membrane-associated
proteins of any type for each given lipid concentration is well-defined
based on experimentally-determined parameters and may be used to draw
conclusions.

\subsubsection*{Simulation setup and basic parameters }

An overview of all parameters used in the reference parameter set
can be found in Table S\ref{tab1:parameter}. The set is based on
the model of a CCP having a diameter of about 100-150 nm and projected
on a two-dimensional plane. The CCP is simulated on a hexagonal simulation
area of about 200 nm diameter, discretized into 217 hexagonal simulation
cells (see Fig. \ref{fig1_model}a), each having a radius of about
9.25 nm and a surface area of 222.5 $nm^{2}$, yielding a total surface
area of 0.064 $\mu m^{2}$. The model includes a regular hexagonal
coat of 54 clathrin molecules, giving rise to a lattice of 162 N-terminal
binding domains. These terminal binding domains are associated with
specific simulation cells - such that a given simulation cell may
have none, one or two terminal domains. We simulate the dynamics for
a total time of 20s, in discrete time steps of 10 $\mu s$ duration
each. The simulation starts with a concentration of 15,000 $\mathrm{PI}(4,5)\mathrm{P}_{2}$
phospholipids per $\mu m^{2}$, uniformly distributed over the simulation
area, resulting in 3 lipid molecules per cell. In the reference parameter
set, the $\mathrm{PI}(4,5)\mathrm{P}_{2}$ concentration remains constant
over the simulation time. The scheme is different for $\mathrm{PI}(3,4)\mathrm{P}_{2}$
: Here the simulation starts with zero lipids. After 10s $\mathrm{PI}(3,4)\mathrm{P}_{2}$
is produced with a rate that elevates the $\mathrm{PI}(3,4)\mathrm{P}_{2}$
concentration within 5 s to the same level as the $\mathrm{PI}(4,5)\mathrm{P}_{2}$
concentration, consistent with expected turnover numbers of PI3Ks\cite{Carpenter1990}
and the copy number\cite{Borner2012} and lifetime \cite{Posor2013}
of PI3K C2$\alpha$ at CCPs. $\mathrm{PI}(3,4)\mathrm{P}_{2}$ lipids
are only produced at clathrin terminal domains, because PI3K C2$\alpha$
binds clathrin at this site\cite{Gaidarov1999}. 

Diffusion is modelled as a jump process between the grid cells, depending
on the molecule\textquoteright{}s diffusion constant and the area
of grid cells of 222.5 $\mathrm{nm}^{2}$. For phospholipid diffusion,
the experimental value of 3 $\mu m^{2}s^{-1}$\cite{Almeida2005}
was employed for both $\mathrm{PI}(4,5)\mathrm{P}_{2}$ and $\mathrm{PI}(3,4)\mathrm{P}_{2}$,
yielding a jump rate of 13482.1 $s^{-1}$. For proteins, it is considered
that diffusion in the cytosol is much faster than lateral membrane
diffusion. Thus, the lateral diffusion constant of membrane-associated
proteins is determined by the diffusion of the bound phosphoinositides,
here taken to be the lipid diffusion constant, divided by the number
of binding sites of the protein. This calculation assumes that the
lipids are sufficiently uncoupled such that their Brownian motion
can be assumed to be uncorrelated \emph{a priori}, but are \emph{a
posteriori} coupled because the protein has to move as a single unit.
Whenever it was attempted to move a protein into a target cell (either
through diffusion, or through binding to the membrane from the cytosol---see
below), availability of sufficient space was tested. The probability
with which such a move would be accepted was set to account for the
actual space left in that particular cell. For example, if a cell
can take four globular proteins of a certain size, and three are already
residing inside the cell, the probability to successfully place a
fourth one is very low because it is unlikely that the space at the
randomly chosen insertion point is completely free of other particles
causing overlaps. We have thus computed placement probabilities for
proteins of different sizes, depending on the space already allocated
on the target cell, and fitted this placement probability with the
sigmoidal function $(1+\exp(ar+b))^{-1}$ with r being the fraction
of the simulation cell area already occupied, and a and b being fitting
parameters (AP-2: a=18.8, b=-2.7; SNX9 \& FCHo2: a=13, b=-2.8; Calm
\& Epsin: a=9.3, b=-3.1). Protein-membrane association parameters
were computed based on published liposome binding experiments. In
these experiments, molecules in the solution bind to a membrane and
eventually bind different kind of lipids. With the model we derived
in the section \textquotedblleft{}Parameterisation of the protein
binding model\textquotedblright{} (see below), we could differentiate
between the two different processes of membrane binding and lipid
binding and could use them separately in our microscopic model. 

For all proteins but CALM, we assumed the macroscopic KD of 7.6 $\mu M$
of AP-2 for binding to $\mathrm{PI}(4,5)\mathrm{P}_{2}$\cite{Jackson2010}.
For CALM, we used 5.8 $\mu M$\cite{Ford2001}. In our binding model,
we assumed CALM and epsin to have one binding site, FCHo2 and SNX9
to have two equal binding sites, and AP-2 to have two strong and two
weak binding sites\cite{Jackson2010}. The on-rate of lipid binding
was set to 10000 $s^{-1}$. The cytosolic protein concentrations were
set such that the protein numbers found under the clathrin coat in
the steady state before the $\mathrm{PI}(3,4)\mathrm{P}_{2}$ increase
matched the copy number of proteins found in CCVs, as described in
a recent quantitative proteomic study\cite{Borner2012}. Several other
parameters (e.g. absolute values of rates) are free parameters that
do not affect the steady states, provided that the parameters above
are chosen as described (table S\ref{tab1:parameter}). In order to
show that the results of the simulation are robust with respect to
parameter changes, the values of important parameters were varied,
yielding a total of 270 different parameter sets that were all tested
as to whether they yielded the same qualitative results. The results
of this parameter sampling are reported in table S\ref{tab2:sampling}
and Figure \ref{fig3:parameterSampling}.

\subsection*{Parametrization of the protein-lipid binding model}

We describe how experimentally-measured lipid binding affinities of
endocytic proteins are transformed into their respective microscopic
simulation parameters. The membrane dissociation constant of proteins
is defined by: 
\[
K_{D}=\frac{[\mathrm{Protein}][\mathrm{Lipid}]}{[\mathrm{Complex}]}
\]
where $[\mathrm{Protein}]$ is the volume concentration of the solvated
protein (e.g. measured as mole per litre), $[\mathrm{Complex}]$ is
the surface concentration of all proteins that are associated to the
membrane (either stably or transiently), and $[\mathrm{Lipid}]$ is
the surface concentration of lipids (e.g. measured in particles per
$nm^{2}$). $K_{D}$ is an experimentally-measurable dissociation
constant. When the lipid and complex concentrations are measured in
the same unit, $K_{D}$ will have the same unit as the protein concentration
(e.g. mole per litre).

In order to convert the macroscopic entity $K_{D}$ into microscopic
simulation parameters, we must specify the dynamical model that the
simulation uses to represent protein binding and dissociation events.
Consider Fig. \ref{fig1_model} c)-e) as an illustration of the binding
dynamics. We consider a simulation cell which is a membrane patch
with area $A_{cell}$. A hypothetical column of cytosol with height
$h$ is considered, yielding a simulation cell volume $V_{cell}$
($h$ and thus $V_{cell}$ can be chosen arbitrarily - they are just
needed as intermediate quantities for the calculation). The volume
concentration of protein is thus given by: 
\[
[\mathrm{Protein}]=\frac{n_{d}}{V_{cell}}
\]
where $n_{d}$ is the average copy number of solvated proteins considered
in the simulation cell volume. We use the convention that all lipid
and associated protein concentrations are measured in particle numbers
per simulation cell area. Thus,
\[
[\mathrm{Lipid}]=n_{lipid}
\]
is the number of lipids per surface area $A_{cell}$. As shown in
Fig. \ref{fig1_model} e), proteins can associate from the cytosol
to state $\mathrm{O}$, in which no phospholipid is bound, and can
dissociate from state $\mathrm{O}$ to the cytosol. The fact that,
depending on the settings of the rates $k_{mb}$ and $k_{mu}$, proteins
can be membrane-associated for some time even if no phospholipid is
present represents the intrinsic membrane affinity of endocytic proteins.
Usually, this intrinsic affinity is small and is controlled by the
ratio $k_{mb}/k_{mu}$. From the associated state $\mathrm{O}$, proteins
may enter stably bound states (including at least one $\mathrm{X}$),
where the identifyer $\mathrm{X}$ specifies which phospholipid binding
sites are filled. We use the convention that a protein cannot dissociate
from the membrane while at least one of the binding sites is filled
with a phospholipid. This is based on the reasoning that pulling a
phospholipid out of the membrane would be energetically extremely
unfavorable. The mechanism of dissociation in our model is thus that
phospholipids must first clear their binding sites, each with rate
$k_{lu}$, before being able to dissociate with rate $k_{mu}$. The
concentration of membrane-associated proteins is the total number
of proteins in any lipid binding states on a surface $A_{cell}$.
For proteins with a single binding site, this is:
\[
[\mathrm{Complex}]=n_{\mathrm{O}}+n_{\mathrm{X}}
\]
where $n_{\mathrm{O}}$ counts membrane-associated proteins with empty
binding sites and $n_{\mathrm{X}}$ counts membrane-associated proteins
with filled binding sites. In proteins with two binding sites, we
have the configurations:
\begin{eqnarray*}
[\mathrm{Complex}] & = & n_{\mathrm{O}\mathrm{O}}+n_{\mathrm{X}\mathrm{O}}+n_{\mathrm{O}\mathrm{X}}+n_{\mathrm{X}\mathrm{X}}\\
 & = & n_{0}+2n_{1}+n_{2}
\end{eqnarray*}
where $n_{k}$ counts the number of membrane-associated proteins with
$k$ binding sites filled. Generally, when a protein has $N$ binding
sites, we have
\[
[\mathrm{Complex}]=\sum_{k=0}^{N}\binom{N}{k}\, n_{k}.
\]
We write down the detailed balance equation for the protein association/dissociation
reactions:
\begin{eqnarray*}
n_{d}k_{mb} & = & n_{0}k_{mu}\\
\frac{n_{d}}{n_{0}} & = & \frac{k_{mu}}{k_{mb}}=:K_{d,mem}
\end{eqnarray*}
where we have defined the microscopic intrinsic membrane affinity
$K_{d,mem}$. The detailed balance equation for the lipid binding/unbinding
reactions are:
\begin{eqnarray*}
n_{k}n_{lipid}k_{lb} & = & n_{k+1}k_{lu}\\
n_{lipid}\frac{n_{k}}{n_{k+1}} & = & \frac{k_{lu}}{k_{lb}}=:K_{d,lip}
\end{eqnarray*}
for all $k=0,...,N-1$, where we have defined the microscopic intrinsic
lipid affinity $K_{d,lip}$. $K_{d,lip}$ varies depending on the
protein and the type of lipid ($\mathrm{PI}(3,4)\mathrm{P}_{2}$ or
$\mathrm{PI}(4,5)\mathrm{P}_{2}$). Combining all previous equations,
we end up with the expression
\begin{equation}
\sum_{k=0}^{N}\binom{N}{k}\left(\frac{n_{lipid}}{K_{d,lip}}\right)^{k}=\frac{K_{d,mem}\, nlipid}{K_{D}\, V_{cell}}\label{eq: finalEquation_N_equal_binding_sites-1}
\end{equation}

This permits the following approach to parametrize the binding model
from experimental measurements of $K_{D}$:
\begin{enumerate}
\item Define a simulation cell height $h$ to be used in the simulations,
arriving at a cell volume $V_{cell}$.
\item Define a microscopic intrinsic protein-membrane affinity by setting
$K_{d,mem}$.
\item Calculate $K_{d,lip}$ by solving Eq. (\ref{eq: finalEquation_N_equal_binding_sites-1})
using the experimental concentration of $n_{lipid}$ and the measured
value of $K_{D}$.
\end{enumerate}
For proteins with one binding site (CALM, Epsin), Eq. (\ref{eq: finalEquation_N_equal_binding_sites-1})
has an explicit solution:

\[
K_{d,lip}^{(1)}=\frac{n_{lipid}}{\frac{n_{lipid}K_{d,mem}}{K_{D,Calm}V_{cell}}-1}.
\]
For proteins with two equal binding sites (SNX9, FCHo$_{2}$), Eq.
(\ref{eq: finalEquation_N_equal_binding_sites-1}) has the following
solution:
\[
K_{d,lip}^{(2)}=\frac{n_{lipid}}{\sqrt{\frac{n_{lipid}K_{d,mem}}{K_{D}V_{cell}}}-1}.
\]
For AP-2, the situation is more complex. Experimentally, there is
evidence that AP-2 has four binding sites. However, these are not
equally strong but rather there are two conformations \cite{Jackson2010}
associated with lipid binding affinities $K_{d,lip}^{+}$ and $K_{d,lip}^{-}$,
respectively. Working out the lipid configuration for this case eventually
yields the equation:
\begin{eqnarray*}
1+2\frac{n_{lipid}}{K_{d,lip}^{+}}+2\frac{n_{lipid}}{K_{d,lip}^{-}}+2\left(\frac{n_{lipid}}{K_{d,lip}^{+}}\right)^{2}+2\left(\frac{n_{lipid}}{K_{d,lip}^{-}}\right)^{2}\\
+8\frac{n_{lipid}^{2}}{K_{d,lip}^{+}\, K_{d,lip}^{-}}+\,10\frac{n_{lipid}^{3}}{(K_{d,lip}^{+})^{2}\, K_{d,lip}^{-}}+10\frac{n_{lipid}^{3}}{K_{d,lip}^{+}\,(K_{d,lip}^{-})^{2}}
\end{eqnarray*}
\vspace{-0.7cm}

\begin{eqnarray}
+\,20\frac{n_{lipid}^{4}}{(K_{d,lip}^{+})^{2}\,(K_{d,lip}^{-})^{2}} & = & \frac{K_{d,mem}\, n_{lipid}}{K_{D}\, V_{cell}}\label{eq_strongWeakFinal}
\end{eqnarray}

To solve this formula for $K_{d,lip}^{+}$ and $K_{d,lip}^{-}$ we
have to use additional information, compared to the equal binding
site model in Equation \ref{eq: finalEquation_N_equal_binding_sites-1},
since we have now one additional unknown which renders our formula
underdetermined without additional experimental input. Fortunately,
in the case of AP-2, such additional input is available: The macroscopic
dissociation constant $K_{D}$ for AP-2 is known for the case where
all four binding sites are operational\cite{Jackson2010}. We will
refer here to that value by $K_{D}^{++--}$. Additionally the value
$K_{D}^{++//}$ is known, the dissociation constant for a mutant where
the two weak binding sites have been mutationally inactivated\cite{Jackson2010}.
Having these two constants, we now can firstly use $K_{D}^{++//}$
in Equation \ref{eq: finalEquation_N_equal_binding_sites-1} based
on two binding sites, which gives us a value for $K_{d,lip}^{+}$
and can then secondly use that value together with $K_{D}^{++--}$
in Equation \ref{eq_strongWeakFinal} to end up with a value for $K_{d,lip}^{-}$.
The numerical values of the binding model parameters used are given
in Table S\ref{tab1:parameter}.

\section*{Results}

\begin{figure*}[!t]
\centering{}\includegraphics[width=0.8\textwidth]{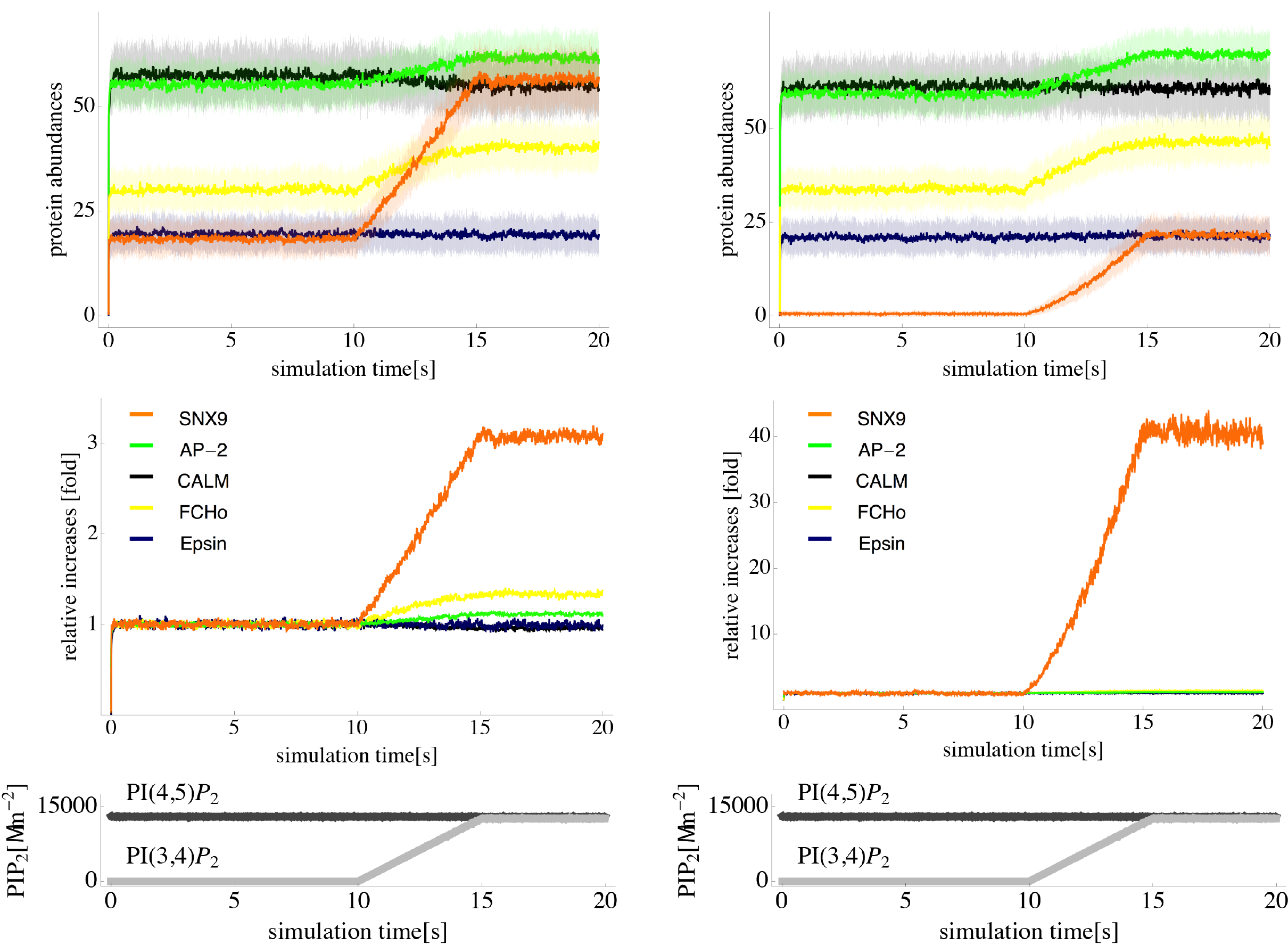}\caption{\label{fig2:results}Spatiotemporal regulation of CME by PI3K C2$\alpha$-mediated
$\mathrm{PI}(3,4)\mathrm{P}_{2}$ synthesis to recruit SNX9. a) Simulated
endocytic protein abundances underneath the clathrin coat of a 2D-projected
CCP over time. Averaged protein copy numbers (solid lines) $\pm$
SD (shaded vertical bars; $n=64$ simulation runs) are shown (top).
Ratio of protein copy numbers normalized to those at time $t=10\, s$
(middle). Concentration profile of $\mathrm{PI}(3,4)\mathrm{P}_{2}$
and $\mathrm{PI}(4,5)\mathrm{P}_{2}$ (bottom). Simulation is divided
in 3 phases: equilibration phase ($0\, s$ - $10\, s$): constant
$\mathrm{PI}(4,5)\mathrm{P}_{2}$ (middle). Concentration profile
of PI(3,4)P2 and PI(4,5)P2 (bottom). Simulation is divided in 3 phases:
equilibration phase ($0\, s$ and zero $\mathrm{PI}(3,4)\mathrm{P}_{2}$
level; rising phase ($10\, s$-$15\, s$): $\mathrm{PI}(3,4)\mathrm{P}_{2}$
is elevated to the level of PIP(4,5)P2; equilibration phase ($15\, s$-$20\, s$).
SNX9 copy numbers are selectively increased in response to $\mathrm{PI}(3,4)\mathrm{P}_{2}$
production to a value sufficient for formation of a ring around the
CCP. Relevance of $\mathrm{PI}(4,5)\mathrm{P}_{2}$ affinity for the
recuitment of SNX9 is tested by a control simulation where SNX9 binds
exclusively $\mathrm{PI}(3,4)\mathrm{P}_{2}$ b). While SNX9 is strongly
recruited to the membrane in both cases (a+b), the scenario in which
SNX9 has no $\mathrm{PI}(4,5)\mathrm{P}_{2}$ affinity (b), does not
reach a sufficient SNX9 level to form a ring around the CCP. This
can be explained by the fact that $\mathrm{PI}(3,4)\mathrm{P}_{2}$
acts as an amplifier that multiplies the available SNX9 concentration
with a certain factor, and thus needs to start from a base level significantly
above zero to reach a substantial target value. Thus, SNX9\textquoteright{}s
affinity to $\mathrm{PI}(4,5)\mathrm{P}_{2}$ is essential. }
\end{figure*}

A representative set of parameters for protein and lipid concentrations,
binding affinities, rates and diffusion constants, was selected (Table
S\ref{tab1:parameter}) and the results are reported in Figure \ref{fig2:results}a.
This data set uses an initially zero $\mathrm{PI}(3,4)\mathrm{P}_{2}$
concentration and a time-independent $\mathrm{PI}(4,5)\mathrm{P}_{2}$
concentration of approximately 15,000 lipids per $\mu m^{2}$as PI
5-kinases are absent from assembled CCPs or coated vesicles\cite{Antonescu2011}.
The time-independent $\mathrm{PI}(4,5)\mathrm{P}_{2}$ concentration
is a conservative assumption as several $\mathrm{PI}(4,5)\mathrm{P}_{2}$-specific
phosphatases have been shown to be present in endocytic CCPs\cite{Perera2006,Nakatsu2010},
potentially resulting in progressive $\mathrm{PI}(4,5)\mathrm{P}_{2}$
hydrolysis. 

With these lipid concentrations, the simulation reaches a steady state
with a strong population of $\mathrm{PI}(4,5)\mathrm{P}_{2}$ binders
such as AP-2, but only few SNX9 molecules associated with CCP membranes
(Fig. \ref{fig2:results}a, top). Based on the experimentally-determined
time course of PI3K C2$\alpha$ recruitment to CCPs\cite{Posor2013}
we analysed the effects of increasing the $\mathrm{PI}(3,4)\mathrm{P}_{2}$
concentration in the simulation with respect to the concentration
of the major endocytic PI binding proteins underneath the clathrin
coat. Strikingly and consistently, SNX9 was the only endocytic protein
which prominently enriched at CCPs concomitant with rising levels
of $\mathrm{PI}(3,4)\mathrm{P}_{2}$, whereas other endocytic PI-binding
proteins such as AP-2, FCHo, CALM, or epsin exhibited comparably minor
changes during the simulated time course of 20 s (Fig. \ref{fig2:results}a). 

\begin{figure*}[!t]
\centering{}\includegraphics[width=1\textwidth]{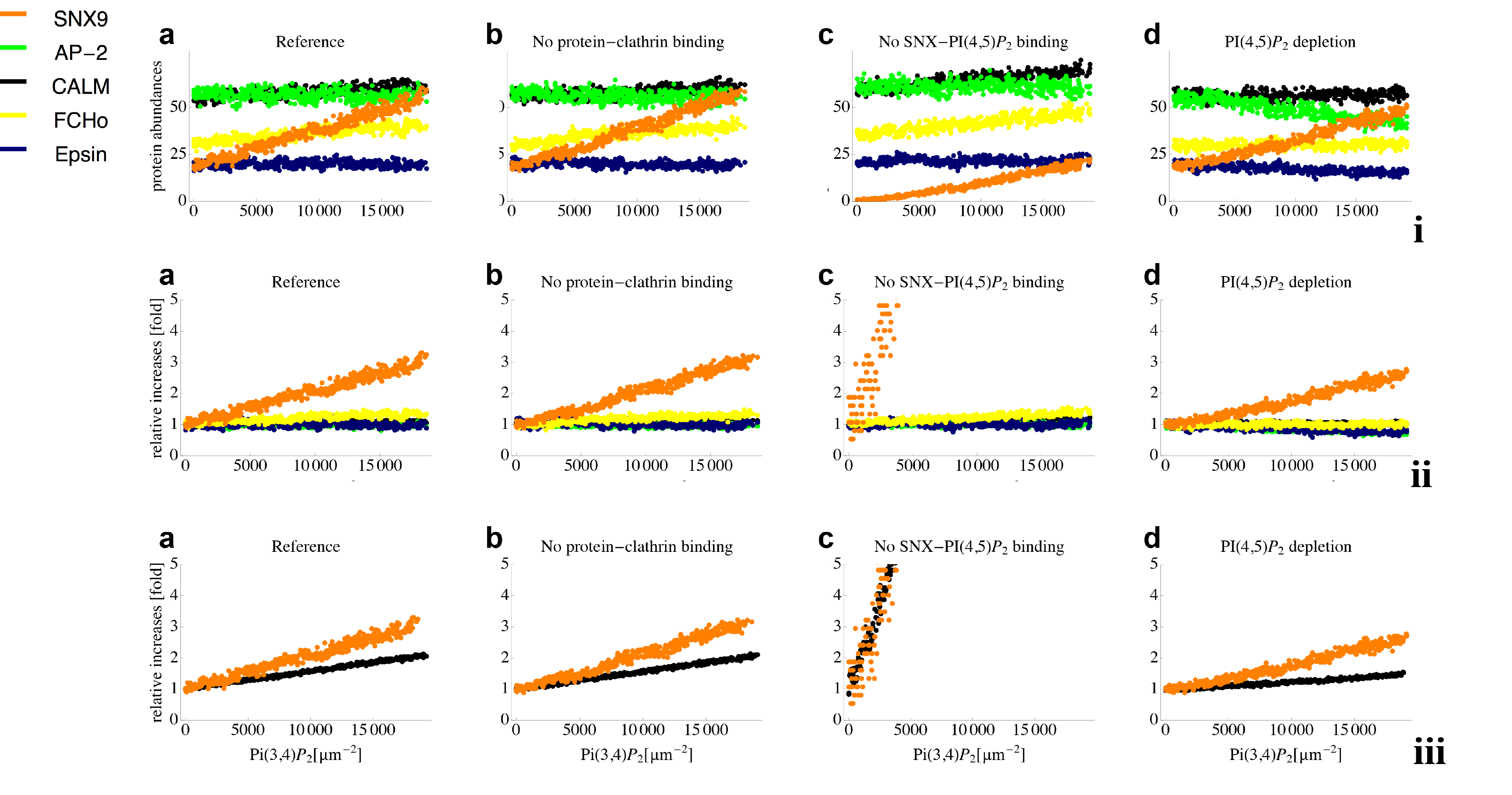}\caption{\label{fig3:parameterSampling}Sensitivity of CCP protein concentrations
on the $\mathrm{PI}(3,4)\mathrm{P}_{2}$ lipids concentration. $\mathrm{PI}(3,4)\mathrm{P}_{2}$
is the main parameter for controlling the SNX9 concentration at the
CCP, while other proteins exhibit little sensitivity on the $\mathrm{PI}(3,4)\mathrm{P}_{2}$
concentration. This can already be anticipated from Fig \ref{fig2:results}a,
where it is seen that the SNX9 population essentially follows the
$\mathrm{PI}(3,4)\mathrm{P}_{2}$ increase, while the other proteins
remain mostly at the same population. To demonstrate this dependency
more clearly, we have taken the simulated time traces of CCP protein
numbers (e.g. Fig \ref{fig2:results}a), and, irrespective of the
actual time, plotted each point of the time trace in a coordinate
system showing the $\mathrm{PI}(3,4)\mathrm{P}_{2}$ concentration
on the x-axis and the protein population on the y-axis. For visual
clarity only data points from simulation times from t=10s to t=15s
were used as the $\mathrm{PI}(3,4)\mathrm{P}_{2}$ concentration was
constant during other times. Row i shows absolute protein numbers,
Rows ii and iii show the relative increase of the protein number by
dividing the absolute number by its value at {[}$\mathrm{PI}(3,4)\mathrm{P}_{2}${]}=0.
The columns a,b,c,d correspond to different choices of simulation
parameters, thereby summarizing the results of 270 simulation settings
with different parameter combinations in a simple way. Four cases
are distinguished here. Compared to the reference parameter set (a),
the other panels have parameter sets that are different from the reference
set by one setting. b) Protein-Clathrin affinity was set to zero.
c) $\mathrm{PI}(4,5)\mathrm{P}_{2}$ affinity to SNX9 was set to zero;
d) The $\mathrm{PI}(4,5)\mathrm{P}_{2}$ concentration is reduced
to half its initial value rather than remaining constant; In all scenarios,
a clear sensitivity of SNX9 to the $\mathrm{PI}(3,4)\mathrm{P}_{2}$
concentration is seen, and the sensitivity of SNX9 population upon
changes in $\mathrm{PI}(3,4)\mathrm{P}_{2}$ is always stronger than
the sensitivity of other proteins to $\mathrm{PI}(3,4)\mathrm{P}_{2}$.
Other proteins show very little sensitive to $\mathrm{PI}(3,4)\mathrm{P}_{2}$.
In row iii, the relative change of SNX9 in the same conditions as
in row ii (orange) is compared to a control simulation (black), where
only SNX9 and no other phospholipid-binding protein was in the system.
It is seen that in cases a, b, and c, SNX9 shows a much stronger response
to changes in $\mathrm{PI}(3,4)\mathrm{P}_{2}$ concentrations when
other proteins are present than when they are not present - the SNX9
number is approximately tripled rather than doubled. This indicates
that $\mathrm{PI}(3,4)\mathrm{P}_{2}$ is not an independent regulator
of endocytosis, but that the regulation depends on the competing interactions
of proteins with the shared lipid pool.}
\end{figure*}

Strong $\mathrm{PI}(3,4)\mathrm{P}_{2}$-dependent SNX9 recruitment
occurs despite the conservative assumption that SNX9 binds to $\mathrm{PI}(3,4)\mathrm{P}_{2}$
and $\mathrm{PI}(4,5)\mathrm{P}_{2}$ equally well (Table \ref{tab1:parameter}b).
This indicates that competition of different proteins for lipids plays
an important role. Other proteins strongly favor $\mathrm{PI}(4,5)\mathrm{P}_{2}$
binding over $\mathrm{PI}(3,4)\mathrm{P}_{2}$ binding, especially
AP-2 and Epsin. These proteins therefore win the competition for $\mathrm{PI}(4,5)\mathrm{P}_{2}$
lipids. This competition in turn acts as an amplification of the $\mathrm{PI}(3,4)\mathrm{P}_{2}$
signal for SNX9 recruitment, which is the only protein that can bind
$\mathrm{PI}(3,4)\mathrm{P}_{2}$ sufficiently well to be sensitive
to this signal. This is indeed confirmed by control simulations in
the absence of competing endocytic proteins resulting in a much weaker
response of the SNX9 population to increased $\mathrm{PI}(3,4)\mathrm{P}_{2}$
levels (see Fig. \ref{fig3:parameterSampling} bottom). 

But why does SNX9 bind both $\mathrm{PI}(3,4)\mathrm{P}_{2}$ and
$\mathrm{PI}(4,5)\mathrm{P}_{2}$ rather than being specific to $\mathrm{PI}(3,4)\mathrm{P}_{2}$
only? To explore this we ran a control simulation in which SNX9 is
assumed to be $\mathrm{PI}(3,4)\mathrm{P}_{2}$-specific with no
more than basal affinity for $\mathrm{PI}(4,5)\mathrm{P}_{2}$. Fig.
\ref{fig2:results}b shows that in this case, $\mathrm{PI}(3,4)\mathrm{P}_{2}$
remains to serve as a strong signal for SNX9 recruitment, providing
a large relative increase of the SNX9 population at the CCP. However,
the absolute copy number of SNX9 molecules at the CCP are comparably
low and remain insufficient for assembly of a closed PX-BAR domain
ring around the neck of nascent late stage CCP. This observation provides
a rational basis for its biochemical behaviour in lipid binding assays
\cite{Posor2013}. Thus, it appears that $\mathrm{PI}(4,5)\mathrm{P}_{2}$
is important for ensuring a basal concentration of SNX9 proteins.
The production of $\mathrm{PI}(3,4)\mathrm{P}_{2}$ in this scenario
``tips the scale'' towards a selective amplification of SNX9 enrichment
to reach a number sufficient to drive the endocytic reaction forward.

To test the robustness of our results, we performed simulations on
varying PI concentrations and binding affinities, yielding a total
of 270 different parameter combinations. In all combinations, increased
$\mathrm{PI}(3,4)\mathrm{P}_{2}$ synthesis induced a robust increase
in the levels of SNX9 (Fig. \ref{fig3:parameterSampling}, Table S\ref{tab2:sampling}).

\section*{Discussion}

The present results suggest a mechanistic interpretation of the experimental
results of Ref. \cite{Posor2013} that is not apparent from the biochemical
data alone. Limited numbers of phospholipids and lipid-binding proteins
interact in a confined space, giving rise to a dynamic and competitive
system with two surprising properties: (1) The activity of $\mathrm{PI}(3,4)\mathrm{P}_{2}$
is amplified as a result of competition of proteins for different
types of lipids. (2) The limited specificity of SNX9, i.e. its ability
to bind to both $\mathrm{PI}(3,4)\mathrm{P}_{2}$ and $\mathrm{PI}(4,5)\mathrm{P}_{2}$
is important for ensuring a sufficient total copy number of SNX9 molecules
at a maturing CCP. The production of $\mathrm{PI}(3,4)\mathrm{P}_{2}$
can therefore be understood as a signal that tips the scales of a
system that is already poised for progression.

Important open questions remain.  In particular it is unclear how
SNX9 recruitment is controlled not only in time but also in space
and whether its enrichment at the CCP neck is compatible with the
presumed localization of PI3k C2$\alpha$ and its lipid product $\mathrm{PI}(3,4)\mathrm{P}_{2}$.
A SNX9-coated neck may act as a molecule sling facilitating the constriction
of the vesicle neck \cite{Shin2008,Yarar2007,Lundmark2009,Ferguson2009}.

\bibliographystyle{naturemag}
\bibliography{Library_yp}

\begin{table*}[!t]
\begin{centering}
\begin{tabular*}{0.64\textwidth}{@{\extracolsep{\fill}}|l|l|l|l|l|l|}
\hline 
a) & SNX9 & AP-2 & CALM & FCHo2 & Epsin\tabularnewline
\hline 
\hline 
copy number\textsuperscript{1} & 7 & 54 & 54 & 33 & 21\tabularnewline
\hline 
footprint {[}nm\texttwosuperior{}{]}\textsuperscript{2} & 34.25 & 84.5 & 8.46 & 83.6 & 9.8\tabularnewline
\hline 
\# lipid bind. sites & 2 & 4 (2s, 2w)\textsuperscript{3} & 1 & 2 & 1\tabularnewline
\hline 
$K_{D,45}${[}$\mu$Mol{]} &  & 7.6 (30.0)\textsuperscript{3,4} & 5.8\textsuperscript{5} &  & \tabularnewline
\hline 
$K_{D,34}$/$K_{D,45}$ & 1.0\textsuperscript{6} & 0.13\textsuperscript{4} & 0.22\textsuperscript{7} & 0.2\textsuperscript{8} & 0.33\textsuperscript{7}\tabularnewline
\hline 
\end{tabular*}
\par\end{centering}

\begin{centering}
\begin{tabular*}{0.64\textwidth}{@{\extracolsep{\fill}}|l|l|l|l|l|l|}
\hline 
b) & SNX9 & AP-2 & CALM & FCHo2 & Epsin\tabularnewline
\hline 
\hline 
$k_{diff}$ {[}$s^{-1}${]} & 6741.57 & 3370.79 & 13483.1 & 6741.57 & 13483.1\tabularnewline
\hline 
$conc{}_{cytosol}$ {[}1/$A_{sim}${]} & 14.2 & 15.6 & 0.33 & 29.5 & 0.09\tabularnewline
\hline 
$K_{a,mem}$ & 0.11 & 5.19 & 5.2 & 0.11 & 4.03\tabularnewline
\hline 
$K_{a,lip45}$ & 3.32 & 0.21 & 9.98 & 3.32 & 9.98\tabularnewline
\hline 
$K_{a,lipweak45}$ &  & 0.0055 &  &  & \tabularnewline
\hline 
$K_{a,lip34}$ & 3.32 & 0.06 & 2.17 & 1.64 & 2.47\tabularnewline
\hline 
$K_{a,lipweak34}$ &  & 0.0574 &  &  & \tabularnewline
\hline 
$K_{a,clathrin}$ & 1.0 & 1.0 & 1.0 & 1.0 & 1.0\tabularnewline
\hline 
\end{tabular*}
\par\end{centering}

\begin{centering}
\begin{tabular*}{0.64\textwidth}{@{\extracolsep{\fill}}|l|l|l|}
\hline 
c) & D {[}$\mu$m\texttwosuperior{}/s{]} (experiment) & $k_{diff}$ {[}$s^{-1}${]} (model)\tabularnewline
\hline 
\hline 
Phospholipids & 3\textsuperscript{9} & 13483.1\tabularnewline
\hline 
\end{tabular*}
\par\end{centering}

\begin{centering}
\begin{tabular*}{0.64\textwidth}{@{\extracolsep{\fill}}|l|l|l|l|l|}
\hline 
d) & time {[}s{]} & dt {[}$\mu$s{]} & cellsize {[}nm\texttwosuperior{}{]} & cellradius {[}nm{]}\tabularnewline
\hline 
\hline 
Global & 20 & 10 & 222.3 & 9.25\tabularnewline
\hline 
\end{tabular*}
\par\end{centering}

\begin{centering}
\begin{tabular*}{0.64\textwidth}{@{\extracolsep{\fill}}|l|l|l|l|}
\hline 
 & k\textsubscript{on,membrane} {[}$s^{-1}${]} & k\textsubscript{on,lipid} {[}$s^{-1}${]} & k\textsubscript{on,clathrin} {[}$s^{-1}${]}\tabularnewline
\hline 
\hline 
Global & 100000 & 10000 & 10\tabularnewline
\hline 
\end{tabular*}
\par\end{centering}

\centering{}%
\begin{tabular*}{0.64\textwidth}{@{\extracolsep{\fill}}|l|l|l|}
\hline 
e) & $k_{kinase}$ {[}$s^{-1}${]} & $k_{phosphatase}$ {[}$s^{-1}${]}\tabularnewline
\hline 
\hline 
Clathrin N-terminal domain & 0.8 & (0.63)\tabularnewline
\hline 
\end{tabular*}\caption{\label{tab1:parameter}Reference parameter set for the reaction-diffusion
simulation. a) Experimentally-determined parameters used for the proteins
simulated in the reaction-diffusion model, references are marked by
superscripts and enlisted below. b) Protein-specific parameters used
in the reaction-diffusion simulation that were either derived from
(a), or computed as described in SI methods. c) Experimental and model
parameters for the two phospholipids. d) Global simulation parameters.
e) Kinase and phosphatase turnover rates. The phosphatase was not
active in the reference parameter set, but only in some alternative
parameter set considered in the parameter sampling (see Table S\ref{tab2:sampling}).
References: 1)\cite{Borner2012}, 2) PDB 2RAI, PDB 2xa7, PDB 1HFA,
PDB 2V0o, PDB 1edu , 3) \cite{Jackson2010}, 4) \cite{Honing2005a}
, 5) \cite{Ford2001}, 6) \cite{Posor2013}, 7) \cite{Itoh2001},
8) \cite{Henne2010a}, 9) \cite{Almeida2005}}
\end{table*}

\begin{table*}[!t]
\begin{centering}
\begin{tabular}{|l|l|l|l|l|}
\hline 
Parameter & {[}SNX9{]} at t=10s & {[}SNX9{]} at t=20s & $\Delta$SNX9 & $\Delta$ others \tabularnewline
\hline 
\hline 
CA = 0 & 18.7 $\pm$ 14.2 & 82.4 $\pm$ 46.4 & 4.4  & 1.02 $\pm$ 0.14 \tabularnewline
\hline 
CA = 1 & 19.4 $\pm$ 13.6 & 82.2 $\pm$ 45.0  & 4.2  & 1.04 $\pm$ 0.14 \tabularnewline
\hline 
CA = 100 & 19.0 $\pm$ 13.1 & 81.9 $\pm$ 43.9 & 4.3  & 1.03 $\pm$ 0.14 \tabularnewline
\hline 
$\mathrm{PI}(4,5)\mathrm{P}_{2}$ Affinity = 0 & 2.6 $\pm$ 2.9 & 57.2 $\pm$ 37.0 & 22.0  & 1.04 $\pm$ 0.15 \tabularnewline
\hline 
$\mathrm{PI}(4,5)\mathrm{P}_{2}$ Affinity = $\frac{1}{2}$ & 21.0 $\pm$ 3.3 & 87.9 $\pm$ 42.2 & 4.2  & 1.02 $\pm$ 0.14 \tabularnewline
\hline 
$\mathrm{PI}(4,5)\mathrm{P}_{2}$ Affinity = 1 & 33.9 $\pm$ 2.3 & 102.3 $\pm$ 42.7  & 3.0  & 1.03 $\pm$ 0.13 \tabularnewline
\hline 
no $\mathrm{PI}(4,5)\mathrm{P}_{2}$ depletion & 19.2 $\pm$ 13.6 & 83.0 $\pm$ 45.1  & 4.3  & 1.15 $\pm$ 0.06 \tabularnewline
\hline 
$\mathrm{PI}(4,5)\mathrm{P}_{2}$ depletion & 19.0 $\pm$ 13.5 & 81.3 $\pm$ 44.7  & 4.3  & 0.92 $\pm$ 0.09 \tabularnewline
\hline 
{[}$\mathrm{PI}(4,5)\mathrm{P}_{2}${]} = 6.500/$\mu$m\texttwosuperior{} & 21.8 $\pm$ 12.0 & 88.9 $\pm$ 45.6  & 4.1  & 1.03 $\pm$ 0.11 \tabularnewline
\hline 
{[}$\mathrm{PI}(4,5)\mathrm{P}_{2}${]} = 13.000/$\mu$m\texttwosuperior{} & 17.2 $\pm$ 13.5 & 81.8 $\pm$ 47.1  & 4.8  & 1.02 $\pm$ 0.15 \tabularnewline
\hline 
{[}$\mathrm{PI}(4,5)\mathrm{P}_{2}${]} = 26.000/$\mu$m\texttwosuperior{} & 18.1 $\pm$ 14.8 & 75.6 $\pm$ 41.3  & 4.2  & 1.06 $\pm$ 0.15 \tabularnewline
\hline 
{[}$\mathrm{PI}(3,4)\mathrm{P}_{2}${]} = $\frac{1}{2}$ {[}$\mathrm{PI}(4,5)\mathrm{P}_{2}${]}  & 19.0 $\pm$ 13.6 & 40.5 $\pm$ 20.7  & 2.1  & 0.95 $\pm$ 0.13 \tabularnewline
\hline 
{[}$\mathrm{PI}(3,4)\mathrm{P}_{2}${]} = {[}$\mathrm{PI}(4,5)\mathrm{P}_{2}${]} & 19.1 $\pm$ 13.9 & 71.5 $\pm$ 20.8  & 3.7  & 1.03 $\pm$ 0.12 \tabularnewline
\hline 
{[}$\mathrm{PI}(3,4)\mathrm{P}_{2}${]} = 2 {[}$\mathrm{PI}(4,5)\mathrm{P}_{2}${]} & 19.0 $\pm$ 13.4 & 134.4 $\pm$ 26.1 & 7.1  & 1.11 $\pm$ 0.11 \tabularnewline
\hline 
\end{tabular}
\par\end{centering}

\caption{\label{tab2:sampling}Overview of the parameter sampling results.
Several uncertain simulation parameters were sampled in order to test
whether our conclusions were robust with respect to parameter changes.
For visual clarity, the 270 results are grouped according to equal
values in individual parameters, each such group is associated with
a row in the table. To ensure sufficient statistics, all 270 possible
combinations of parameter values were simulated for 8 runs. The variation
of the other parameters is described in terms of the standard deviation
over all samples reported as a second number in table entries of columns
2,3, and 5. In the first column all sampled parameter values are shown.
CA - The affinity of all proteins to bind the N-terminal domain of
clathrin. $\mathrm{PI}(4,5)\mathrm{P}_{2}$ Affinity - The affinity
of SNX9 to $\mathrm{PI}(4,5)\mathrm{P}_{2}$ in relation to $\mathrm{PI}(3,4)\mathrm{P}_{2}$.
$\mathrm{PI}(4,5)\mathrm{P}_{2}$ depletion - the concentration of
the lipid is decreased to half its starting value between t=10s and
t=15s (or not). {[}$\mathrm{PI}(4,5)\mathrm{P}_{2}${]} - concentration
of the lipid in the simulation. {[}$\mathrm{PI}(3,4)\mathrm{P}_{2}$
= x{[}$\mathrm{PI}(4,5)\mathrm{P}_{2}${]} - the target concentration
of the lipid $\mathrm{PI}(3,4)\mathrm{P}_{2}$ in relation to the
initial concentration of $\mathrm{PI}(4,5)\mathrm{P}_{2}$. The parameter
values of the reference parameter set are highlighted in green. The
parameters were assessed based on the properties listed in the first
row, columns 2-5. In all scenarios, the concentration of SNX9 at t=10s
(before addition of $\mathrm{PI}(3,4)\mathrm{P}_{2}$) was below the
putative threshold of 40 to 50 copies required for ring formation
(column 2). The only parameter with a significant influence on this
quantity is the $\mathrm{PI}(4,5)\mathrm{P}_{2}$ affinity of SNX9.
On the other hand (column 3) the concentration of SNX9 at t=20s increases
substantially in all cases, and is usually well above 50. Higher $\mathrm{PI}(4,5)\mathrm{P}_{2}$
affinities and higher $\mathrm{PI}(3,4)\mathrm{P}_{2}$ concentrations
have the most prominent effect on the final SNX9 concentration. While
a certain SNX9 affinity to $\mathrm{PI}(4,5)\mathrm{P}_{2}$ is needed
to ensure a baseline concentration of SNX9, the number of recruited
SNX9 copies mainly depends on the amount of $\mathrm{PI}(3,4)\mathrm{P}_{2}$
produced (see also Fig \ref{fig3:parameterSampling}). Column 4 reports
the relative increase of SNX9 when dividing the SNX9 copy number after
$\mathrm{PI}(3,4)\mathrm{P}_{2}$ production by the SNX9 copy number
before the increase of $\mathrm{PI}(3,4)\mathrm{P}_{2}$. Low $\mathrm{PI}(4,5)\mathrm{P}_{2}$
concentrations also results in a high relative increase of SNX9, but
are however starting from low initial SNX9 copy numbers and mostly
do not reach sufficient total SNX9 copy numbers. Consequently, both
the ability to bind to $\mathrm{PI}(4,5)\mathrm{P}_{2}$ and sufficient
$\mathrm{PI}(3,4)\mathrm{P}_{2}$ levels are necessary to accumulate
sufficient SNX9. Column 5 shows the relative change in copy numbers
of the other proteins between t=10s and t=20s.}
\end{table*}

\clearpage
\end{document}